# Mining Event Logs to Support Workflow Resource Allocation

Tingyu Liu, Yalong Cheng, Zhonghua Ni*

School of Mechanical Engineering, Southeast University, Nanjing 210096, China

Jiangsu Key Laboratory for Design and Manufacture of Micro-Nano Biomedical Instruments, Southeast University, Nanjing 210096, China

E-mail addresses: tyliu@live.com (T. Liu), chengyalong.seu@gmail.com (Y. Cheng), nzh2003@seu.edu.cn (Z. Ni)

**Abstract:** Currently, workflow technology is widely used to facilitate the working process in enterprise information systems (EIS), and it has the potential to reduce design time, enhance product quality and decrease product cost. However, significant limitations still exist: as an important task in the context of workflow, many present resource allocation operations are still performed manually, which are time-consuming. This paper presents a data mining approach to address the resource allocation problem (RAP) and improve the productivity of workflow resource management. Specifically, an Apriori-like algorithm is used to find the frequent patterns from the event log, and association rules are generated according to predefined resource allocation constraints. Subsequently, a correlation measure named lift is utilized to annotate the negatively correlated resource allocation rules for resource reservation. Finally, the rules are ranked using the confidence measures as resource allocation rules. Comparative experiments are performed using C4.5, SVM, ID3, Naïve Bayes and the presented approach, and the results show that the presented approach is effective in both accuracy and candidate resource recommendations.

## 1. Introduction

Workflow is now an embedded technology in many enterprise information systems (EIS, e.g. PLM, ERP, CRM, SCM and B2B applications etc.). Workflow resource allocation serves as an indispensable link between workflow activities and resources, and it directly determines the execution quality of the workflow activities [1-3].

Based on our investigation, most of the resource allocation tasks in present workflow management systems are usually performed using a role-based approach [2, 4, 5]. That is, to divide the workflow resources (actors) into different candidate groups based on their role and the organization properties. Once the workflow cases are originated, the workflow engine assigns the works to proper resource groups [4, 6]. Such resource allocation is somewhat coarse-grained and may fail in some situations. For example, in the manufacturing enterprises, a manufacturing process sheet work might be predefined to be undertaken by the resources with the role "process planning designer". Actually, some of the processes planning works have to be further assigned to a smaller group of one or more qualified designers instead of all the process planning designers. Thus, the present resource allocation methods may make inappropriate staff assignments and the final quality of the products may suffer from it. Therefore, in some industries such as the manufacturing enterprises, most of run-time workflow resource allocation works are still performed manually by the administrators. The number of administrators is usually small, whereas the activities are of great abundance in some cases. That makes it a time-consuming work to allocate the workflow resources manually.

Fortunately enough, contemporary workflow applications usually record the business events in event logs. These logs typically contain information about events referring to a case, an activity, and an originator [7-10]. The case (also referred to as process instance) is a work that is being handled, e.g. a process planning sheet design, a compressor design, an NC programming, etc. As the atomic element of the case, an activity is an instance of a workflow task. An originator is a resource (usually a person) that executes the activity[6]. In this paper, a Process refers to a workflow template of the case, a Task represents a series of similar activities, and a Resource refers to a task performer.

This paper presents an Apriori-like algorithm [11, 12] to find frequent patterns from the workflow logs, which are used to generate rules according to a "resource allocation rule constraint". All the negative correlated rules are annotated with a rule evaluation measure referred to as "correlation measures". Then, the selected rules are ranked in a descending sequence by their confidence, and the final rules are then recommended to workflow administrators at workflow run-time.

The major contributions of this paper are as follows: First, it designs a closed-loop workflow framework for a more intelligent and finer-grained resource management. Second, it proposes an association rule mining approach to find the logics between workflow resources and the activities, which would help decision-making in resource allocation.

The remainder of this paper is presented as follows: In Section 2, we design a closed-loop workflow architecture for optimizing resource allocation. Later on, we study the workflow event models and their relationships in Section 3, and then propose our mining approach in Section 4. In Section 5, we empirically compare some classification algorithms (C4.5, SVM, ID3, and Naïve Bayes) with our approach. In Section 6, we discuss some possible improvements. Finally, we discuss the related works in workflow resource allocation in Section 7, and conclude this paper in Section 8.

## 2. A closed-loop workflow framework for resource allocation

Our work is based on a National Defense Project named Agile Process Preparation System (APPS) for a large radar-manufacturing corporation [13] in Nanjing, Jiangsu, China. APPS is a process-aware information system, and it applies a workflow module to manage the works of CAX units (e.g. CAD, CAM, CAPP, etc.). This workflow module manages the resources (performers/actors) using a closed-loop approach. The framework of the approach is illustrated in Fig. 1.

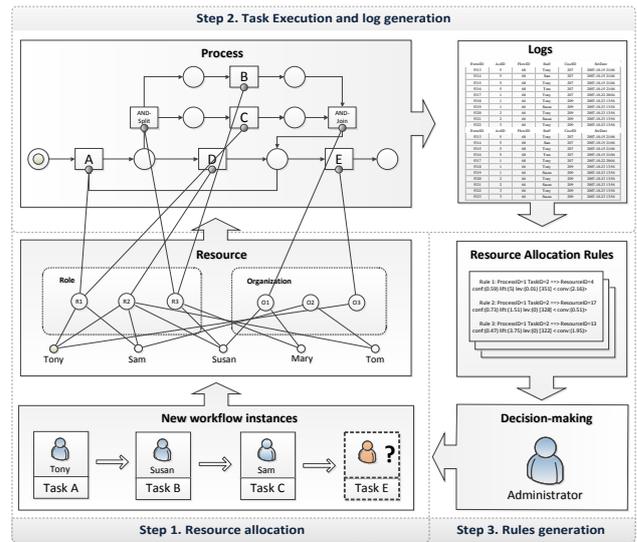

**Fig. 1** Overview of the approach

The closed-loop workflow resource allocation approach mainly includes three steps:

Step 1: The execution history of the workflow activities is recorded in a transaction log referred to as workflow log.

Step 2: The system utilizes an Apriori-like association rule mining algorithm to extract the resource allocation knowledge from the workflow log.

Step 3: Once a new workflow activity is originated, the system automatically recommends the administrator with a default resource and other proper candidates according to the mined association rules. The workflow administrator may simply approve the default assignment or choose another resource in the candidate list for the work considering the reality.

## 3. From workflow log to Resource Allocation Rules

Our goal is to distill resource allocation rules with high prediction accuracy out of the workflow event log. A workflow event typically includes three primary kinds of information: the workflow process information, the workflow task information and the resource information. The association rule involving these three dimensions without repeated predicates falls in the multi-dimensional association rules mining domain[14].

### 3.1. Models and entities in workflow

#### 3.1.1. Workflow model

To illustrate, we use a product planning process $p_{66}$, and the workflow diagram consistent to the process is depicted in Fig. 2.

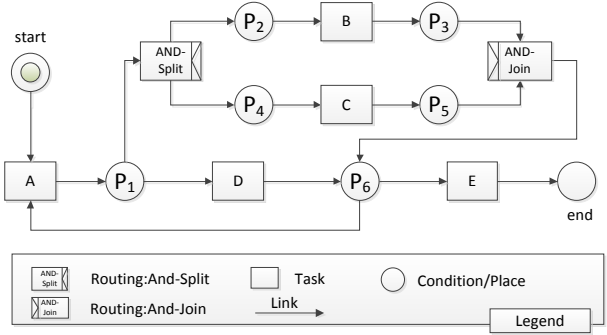

Fig. 2 A sample process $p_{66}$ modeled using WF-NET

Fig. 2 shows a simplified workflow process $p_{66}$ modeled with WF-NET[4]. This process includes five tasks (A, B, C, D, and E), a parallel routing (the AND-Split and the AND-Join), and two selective routings (OR-Split $P_1$ and OR-Join $P_6$).

**Table 1** A Part of the Workflow Log

| EventID | ActID | FlowID | Staff | CaseID | SetDate |
|---|---|---|---|---|---|
| 5313 | 1 | 66 | Tony | 203 | 2007-10-15 21:06 |
| 5314 | 1 | 66 | Sam | 204 | 2007-10-15 21:09 |
| 5315 | 4 | 66 | Mary | 203 | 2007-10-15 21:10 |
| 5316 | 1 | 66 | Sam | 205 | 2007-10-15 21:11 |
| 5317 | 1 | 66 | Tony | 205 | 2007-10-15 21:13 |
| 5318 | 1 | 66 | Tony | 206 | 2007-10-16 13:46 |
| 5319 | 2 | 66 | Tom | 204 | 2007-10-16 13:47 |
| 5320 | 1 | 66 | Sam | 203 | 2007-10-16 13:49 |
| 5321 | 4 | 66 | Susan | 203 | 2007-10-16 13:50 |
| 5322 | 1 | 66 | Mary | 206 | 2007-10-16 13:51 |
| 5323 | 2 | 66 | Mary | 204 | 2007-10-16 13:52 |
| 5324 | 3 | 66 | Tony | 204 | 2007-10-16 13:53 |
| 5325 | 3 | 66 | Tom | 204 | 2007-10-16 13:54 |
| 5326 | 1 | 66 | Sam | 206 | 2007-10-16 13:55 |
| 5327 | 2 | 66 | Tom | 205 | 2007-10-16 13:56 |
| 5328 | 5 | 66 | Susan | 203 | 2007-10-16 14:02 |
| 5329 | 4 | 66 | Sam | 206 | 2007-10-23 14:04 |
| 5330 | 2 | 66 | Mary | 205 | 2007-10-23 14:05 |
| 5331 | 1 | 66 | Susan | 204 | 2007-10-23 14:06 |
| 5332 | 1 | 66 | Mary | 206 | 2007-10-23 14:08 |
| 5333 | 3 | 66 | Sam | 205 | 2007-10-23 14:10 |
| 5334 | 3 | 66 | Susan | 205 | 2007-10-23 14:13 |
| 5335 | 2 | 66 | Tony | 206 | 2007-10-23 14:14 |
| 5336 | 3 | 66 | Susan | 206 | 2007-10-23 15:31 |
| 5337 | 3 | 66 | Tom | 206 | 2007-10-23 15:33 |
| 5338 | 4 | 66 | Sam | 204 | 2007-10-23 15:37 |
| 5339 | 3 | 66 | Susan | 206 | 2007-10-23 15:38 |
| 5340 | 5 | 66 | Tom | 205 | 2007-10-23 15:42 |
| 5341 | 5 | 66 | Susan | 205 | 2007-10-23 15:43 |
| 5342 | 5 | 66 | Tom | 204 | 2007-10-23 15:44 |
| 5343 | 5 | 66 | Susan | 206 | 2007-10-23 15:47 |
| 5344 | 5 | 66 | Tom | 205 | 2007-10-23 15:50 |

Table 1 is an event log sample consistent with the process $p_{66}$. This sample mainly includes some entities of WfMS: resource and task, case, and process, the "EventID" is the identity of the log and the "CaseID" referred to the identity of the instances of $p_{66}$.

This process is modeled in a Petri-net-like model referred to as WF-NET. Entities in this diagram are: process, task, and routing, etc [4, 15]. First, in Task A, the system automatically searches the database for similar cases. If there are cases meeting the requirements, the process would be submitted to Task D, and the designer will download the case documents and alter them. If there is no well-suited case, the work would be passed through a parallel routing to both Task B and Task C, and new design tasks will be assigned to corresponding designers for Task B and co-designer for Task C. When both the Task B and C are finished, the designed document will be archived in Task E and the whole design process ends. Note that, we do not consider iteration routings, like the directed line from place $P_6$ to Task A.

A workflow log generates as the works transact from one step to another consistent with the control flow of the process in Fig. 2. Note that we filter out some notions such as time stamps, event types, which are not helpful here.

Neither do we consider the ordering of the events corresponding to different cases.

We consider workflow logs as a sequence of distinct workflow activities, where subsequences, such as the cases (a case is an instance of a process.), can be observed by usually long gaps between consecutive queries. For example, as is shown in Table 1, assume that a workflow log consists the following workflow transaction events: $e_1:(p_1,t_2,r_4)$, $e_2:(p_2,t_1,r_3)$, $e_3:(p_1,t_2,r_9)$, $e_4:(p_2,t_1,r_7)$, $e_5:(p_3,t_3,r_{10})$. This sequence can be divided into activity set view according to process id and task id: Activity Set 1 ($p_1,t_2$): ($e_1$, $e_3$) ; Activity Set 2 ($p_2,t_1$): ($e_2$, $e_4$); Activity Set 3 ($p_3,t_3$): ($e_5$), where each Activity corresponds to a same pair of process and task. Thus, we get a schematic view of the sample of the workflow log in Table 2:

**Table 2** A schematic view of the workflow log

| Case ID | Log events |
|---|---|
| 1 | (A, Tony), (B, Susan), (C, Tom), (E, Mary) |
| 2 | (A, Tony), (D, Jim), (E, Mary) |
| 3 | (A, Tony), (B, Susan), (C, Tom), (E, Mary) |
| 4 | (A, Tony), (D, Jim), (E, Mary) |
| 5 | (A, Tony), (B, Sam), (C, Sam), (E, Tony) |
| 6 | (A, Jim), (B, Susan), (C, Sam), (E, Tony) |

For convenience, we define some models used in the mining process by adopting some notions defined in Ref. [15].

**Definition 1. (Workflow Process)**

A process indicates the working tasks and the orders in which this should be done. Let $P=\{p_1,p_2,...,p_{n_P}\}$ be a set of processes, where $p_i$ is a process. A task is an atomic unit of a process, let T be a set of tasks $T_i=T(p_i)=\{t_1,t_2,...,t_{n_i}\}$, $i\in\{1,2,...,n_P\}$. Let $R$ be the set of performers/originators (i.e., staffs, resources, or agents), $R=\{r_1,r_2,...,r_{n_r}\}$.

Here, $n_P=|P|$, and $n_i=|T(p_i)|$, and $T(p_i)$ is the task set of the process $p_i$. $n_r$ is the number of the resource units, $n_r=|R|$.

Fig. 2 shows a simplified workflow process. In the theory of workflow modeling [4, 16], routing determines the control flow of the process, the order of the performances of the tasks, and is usually performed automatically. In Fig. 2, there are 5 tasks need to be assigned manually in $p_{66}$, $TasksIn\,\text{Pr}ocess(p_{66})=\{t_1,t_2,t_3,t_4,t_5\}=\{A,B,C,D,E\}$.

The task and process entities fall in the control flow aspect of WfMS, and the performer/originator entities fall in the organizational model prospect. There are some other entities in the organizational prospect like the organizational unit, the organization and the role. However, we do not take into account such information in this paper.

### 3.1.2. Workflow log

To handle cases is the primary objective of a workflow system, where the tasks of similar cases are organized in the same ways, namely workflow processes. In other words, a case is an instance of some process. The transaction events generate as the cases run in the workflow system. A workflow log is a collection of transaction events of the task executions. Workflow log records the information such that: (1) each event refers to an activity, which is a task in the process, (2) each event refers to a case, which is an instance of process, (3) each event refers to a performer, a workflow resource (probably a staff, or just a printer) executing the activity. Therefore, we abstract the event log as a set of quadruples:

($case, task, resource, timestamp$)

The tuple indicates that a resource executes the task of an instance of process at a specific time. In this paper, we focus on who execute what task in which process and do not care much about the execution time and the sequence of the tasks (cases). What interests us is thus the co-occurrence of processes, tasks, and resources. As a case is an instance of a process, we can easily get the process identities with case identities. Let's denote the sets $P = \{p_1, p_2, ..., p_K\}$, $T = \{t_1, t_2, ..., t_M\}$, $R = \{r_1, r_2, ..., r_N\}$ to be the sets of K processes, M tasks and N resources in log $L$. With the log data pretreatment (omit the time information and get the process information from the cases), we can translate each quadruple to a triple of $(process, task, resource)$.

Consider a sample of the workflow transaction log in our workflow management system shown in Table 1. Typically, the log contains thousands of records, where each record refers to a certain workflow activity. An activity is an execution composed of a process, a task, and a resource, probably a staff.

**Definition 2. (Event log)**

Let $E = P \times T \times R$ be the set of (possible) events, an event $e_s : (p_i, t_j, r_k)$ logs a workflow activity comprised of a process $p_i$, a task $t_j$ and a resource $r_k$ (the originator). $C = E*$ is the set of possible event sequences (traces describing a case). $L \in B(C)$ is an event log, here $B(C)$ is the set of all bags (multi-sets) over C. Each element of $L$ denotes a case.

It means that the task $t$ of process $p$ is executed by resource $r$. In WfMS, $R = \{r_1, r_2, ..., r_{n_r}\}$, and $n_r$ is the number of the resource units, $n_r = |R|$.

### 3.2. Resource allocation rules representation

In this work, there is a multidimensional data warehouse with four interrelated relations as is shown below:

- Workflow_log (EventID, ProcessID, TaskID, ResourceID, EventType, CaseID),
- Process (ProcessID, ProcessName, ),
- Task (TaskID, TaskName, ProcessID, TaskType, Desription, ),
- Resource (ResourceID, ResourceName, HasOrgEntity, HasRoleEntity),

Where Process, Task, Resource are three dimension tables. These tables are linked to the Workflow_log table via three keys: ProcessID, TaskID and ResourceID. The correlated star schema of our warehouse is depicted in Fig. 3.

The log for mining must meet the requirements brought forth in [15] before process mining. The log data is preprocessed to get qualified for process mining, these pre-processing include revising or eliminating the faulty or unsound data. Faulty or unsound data mainly refer to those incomplete data, which are short of the necessary data items such as the activity, process, case or originator. Therefore, before mining, some preparations are necessary. First, if the log data comes from different data sources, then corresponding translation is essential for a unified data format and easier to process in the following steps, these works are some data processing methods such as coding, simplification, etc.

In the workflow log in Table 1, the EventID, the ActID and the CaseID are the # of the workflow events, the workflow activities and the workflow process cases, respectively. The Resource is the name of the originators. The log segment includes 33 events, involves 5 activities, 5 resources, 4 cases and a process $p_{66}$.

In this paper, the attributes in the workflow log are all nominal. Instead of searching on only one attribute like process, we need to run through multidimensional attributes including process, task and resource, treating each attribute-value pair as an itemset. We use the multidimensional data shown in the star schema in Fig. 3 to construct a data cube [17-19]. The generalization of group by, roll-up and cross-tab ideas is to aggregate the dimensions. In Fig. 3, it is a 3-dimensional data cube, and the traditional GROUP BY generates the 3-dimensional data cube core. The lower-dimensional aggregates appear as points, lines, planes, or cubes hanging off the data cube core.

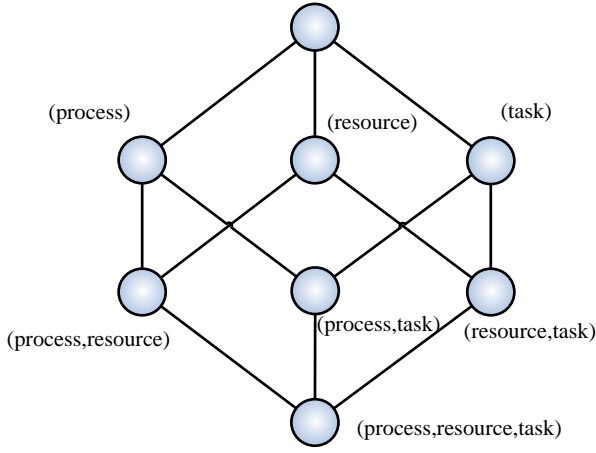

**Fig. 3** The 3-D data cube[14]

Data cubes are well suited for mining multidimensional association rules. Fig. 3 shows the lattice of cuboids defining a data cube for the dimensions process, task and resource. An association rule has the form like $LHS \Rightarrow RHS$, that is, from Left Hand Side (LHS) to Right Hand Side (RHS). By using the data cube, we may get several different multidimensional rules.

Let us now see an example of a single frequent 3-itemset $F_3 : \{p_1, t_1, r_9\}$, which is derived from the event log using the algorithm in section 4.1. The nonempty subsets of $F_3$ are $\{p_1, t_1\}$, $\{t_1, r_9\}$, $\{p_1, r_9\}$, $\{p_1\}$, $\{t_1\}$, $\{r_9\}$. Thus, we can get the association rule in different forms:

$p_1 \wedge t_1 \Rightarrow r_9$,   $p_1 \Rightarrow t_1 \wedge r_9$,   $p_1 \Rightarrow r_9$,   $t_1 \Rightarrow p_1$

$p_1 \wedge r_9 \Rightarrow t_1$,   $r_9 \Rightarrow p_1 \wedge t_1$,   $t_1 \Rightarrow r_9$,   $r_9 \Rightarrow t_1$

$t_1 \wedge r_9 \Rightarrow p_1$,   $t_1 \Rightarrow r_9 \wedge p_1$,   $p_1 \Rightarrow t_1$,   $r_9 \Rightarrow p_1$

Whereas some of the rules are of no help to resource allocations, e.g., the rules in the form of $t_1 \Rightarrow p_1$ means that the task $t_1$ of process $p_1$ is frequent performed in the system. Hence, we have to use the dimension/level constraints[20] to filter out the rules with little interest.

**Definition 3. (Resource Allocation Rule Constraint)**

For an activity of Task Y in Process X, and the Resource Z, our research objective in this paper is to find the resource allocation rule as follows called PTR (process, task to resource) metarule:

$$process(X, "1...n_p") \wedge task(Y, "1...n_{t_{p_X}}") \Rightarrow resource(Z, "Terry...Mary") \quad (1)$$

If we filter the rules using constraints in Definition 3, then only the rule $p_1 \wedge t_1 \Rightarrow r_9$ is qualified output. We can get a list of the rules by iterating this step to all of the frequent itemsets in the event log. "Find the execution of what task may promote the working frequency of the resources in the same case (the instance of a process)" is an association rules mining query, which can be expressed in a data mining query language (DMQL) as follows:

Mine multi-dimensional association rules as
    Process+(A, A.ProcessName) and
    Task+(B, B.TaskName, ?[C], _, _) $\Rightarrow$ Resource+(C, C.ResourceName, ?[D], ?[E])
    from Workflow_log
Where P.ProcessID=A
Group by A, B
Having
With minimum support=*min_sup* and minimum confidence=*min_conf* and minimum lift=*min_lift*

This mining query allow the generation of association rules in the form as below:

Process (9, "Process Planning File Design") and Task (1, "Search for similar file", 9, "others", "Search for similar file template in the file database.") $\Rightarrow$ Resource(3, Tom, "1,2,5", "2,6,7,9") [conf:(0.59); sup:(0.032); lift:(5)]

The rules mean that if a work activity of the Task 1 in Process 9 is to be executed, there is a 59% probability that the work will be performed by Resource 3, Tom. A further indication of the rule is that, 3.2% of all the workflow events fulfilled all the criteria, and the lift measure of this rule is 5, indicating it a positively correlated association rule.

**4. Generate, annotate and rank: A three-stage**

### approach to mine resource allocation rules

We have introduced some basic terms in resource allocation rules mining in previous subsections. In order to mine the multidimensional association rules from the event logs, in this subsection, we present a three-stage approach to get the useful rules.

**Stage 1.** Generated raw resource allocation rules: Find all frequent 3-itemsets, and generate resource allocation rules using the rule constraint in Definition 3, and by definition, each of the itemsets should satisfy the minimum support and minimum confidence.

**Stage 2.** Annotate the rules by Negative Correlation Annotation algorithm.

**Stage 3.** Make a rule sequence by confidence of the rules using resource allocation rules sorting method.

### 4.1. Frequent resource allocation rules generation

In association rules mining domain, an itemset $I$ is frequent only if its *support* value satisfies the minimum support threshold $min\_sup$. The term *support* here is also referred to as *relative support*, and it indicates the occurrence frequency of the itemset[14]. For association rules mining in the form of $p \wedge t \Rightarrow r$, we define the term support as:

$$\sup(p \wedge t \Rightarrow r) = \frac{count(p,t,r)}{count(L)} \quad (2)$$

As is shown in Eq.(2), the support measure is the percentage of transactions in $L$ that contain the itemset $(p,t,r)$, The function $count(L)$ returns the number of records in the log, and the $count(p,t,s)$ returns the count of event logs corresponding to the process $p$, task $t$ and resource $r$.

Frequent itemset mining leads to the discovery of associations and correlations among items in large transactional data sets. However, this can be a time-consuming procedure. In this paper, we use the Apriori algorithm to find frequent patterns. Apriori is a classical algorithm proposed by R. Agrawal and R. Srikant in 1994 for mining association rules, and is proved to be efficient and scalable for both artificial and real world data sets[11, 12]. The high performance of this algorithm is based on the priori knowledge that all nonempty subsets of a frequent itemset must also be frequent, and here we use its contraposition.

We apply the Apriori algorithm along with the "Resource Allocation Constraint". According to Definition 3, the frequent itemset should be 3-dimentional, and the frequent itemsets must satisfy $min\_sup$ threshold. We can get the mining algorithm below:

**Mining Multidimensional association rules[14] from workflow event logs.**

**Algorithm: Frequent-pattern generation.** Find frequent itemsets using an iterative level-wise approach based on Apriori candidate generation.

**Input:**
- $L$, the workflow event log;
- $min\_sup$, the minimum support count threshold.

**Output:** $F_3$, frequent 3-itemsets in $L$.

**Method:**

(1) $F_1 = find\_frequent\_1-itemsets(L)$;

(2) **for** $(k=2; F_{k-1} \neq \phi; k++)\{$

(3) $\quad C_k = apriori\_gen(F_{k-1})$

(4) $\quad$ **for each** transaction $t \in L$ {//scan L for counts

(5) $\quad\quad C_t = subset(C_k, t)$ //get the subsets of t that are candidates

(6) $\quad\quad$ **for each** candidate $c \in C_t$

(7) $\quad\quad\quad c.count++$;

(8) $\quad$ }

(9) $\quad F_k = \{c \in C_k | c.count \geq min\_s$

(10) }

(11) **return** $L = \bigcup_3 L_3$; //Generate frequent 3-itemsets with dimension constraints.

Procedure $apriori\_gen(F_{k-1}; frequent(k-1)-itemsets)$

(1) **for each** itemset $l_1 \in F_{k-1}$

(2) $\quad$ **for each** itemset $l_2 \in F_{k-1}$

(3) $\quad\quad$ **if** (

(4) $\quad (l_1[1] \in l_2[1]) \wedge (l_1[2] \in l_2[2]) \wedge ...$
$(l_1[k-2] \in l_2[k-2]) \wedge ((l_1[k-1] \in l_2[k-1])$

(5) $\quad )$

(6) $\quad$ **then** {

(7) $\quad\quad c = l_1 \otimes l_2$; //joint step: generate candidates

(8) $\quad\quad$ **if** has_infrequent_subset($c$, $F_{k-1}$) **then**

(9) $\quad\quad\quad$ **delete** $c$; //prune step: remove unfruitful candidate

(10) $\quad\quad$ **else add** $c$ to $C_k$;

(11) $\quad$ }

(12) **return** $C_k$;

Procedure *has_infrequent_subset* ( $c$ : *candidate k−itemset*; $F_{k-1}$ : *frequent(k−1)−itemsets* ); // use the prior knowledge

(1) **for each** $(k-1)$-subset s of $c$

(2) $\quad$ **if** $s \notin F_{k-1}$ **then**

(3) $\quad\quad$ **return** TRUE;

(4) **return** FALSE;

Once that the frequent 3-itemsets from the log have been found, it is straightforward to generate strong rules from them. *Strong rules* are those who both satisfy *minimum support threshold* (*min_sup*) and *minimum confidence threshold* (*min_conf*). The rule $p \wedge t \Rightarrow r$ has *confidence c* in the transactions log set L, where *c* is the percentage of transactions in L containing $p \wedge t$ that also contain $r$. It is a conditional probability:

$$confidence(p \wedge t \Rightarrow r) = P(r \mid p \wedge t)$$
$$= \frac{support\_count(p \wedge t \wedge r)}{support\_count(p \wedge t)} \quad (3)$$

To convert the frequent itemsets into strong resource allocation rules, we use the constraint in Definition 3 to confine the dimension and form of the mined rules, and we may get a list of these "qualified in form" rules below:

Rule 1:     process=8 task=1 655 ==> resource=19 655    conf:(1)
Rule 2:     process=7 task=1 206 ==> resource=17 199    conf:(0.97)
Rule 3:     process=5 task=8 296 ==> resource=4 276 conf:(0.93)

4.2. Negatively correlated rules annotation

In the previous sections, we discussed the method of finding the frequent executors for the workflow tasks. However, the rules mined with the support-confidence framework discussed above may disclose some not so interesting event relationships[21, 22]. Let us examine the attached rules:

Rule 1: ProcessID=1 TaskID=2 ==> ResourceID=4    conf:(0.59) [support_count=967]
Rule 2: ProcessID=1 TaskID=2 ==> ResourceID=17    conf:(0.20) [support_count=328]
Rule 3: ProcessID=1 TaskID=2 ==> ResourceID=13    conf:(0.13) [support_count=213]

As illustrated in the list, all of the rules are above the support/confidence threshold. However, Rule 2 would be misleading when $P(p_1 \wedge t_2 \Rightarrow r_{17}) = 0.20 < P(r_{17}) = 0.40$. Therefore, by definition, LHS($p_1 \wedge t_2$) and RHS($r_{17}$) are actually negatively correlated as the existence of LHS actually decreases the likelihood of RHS.

Furthermore, in the resource allocation rules mined from the workflow logs (in the form of $p \wedge t \Rightarrow r$),

$$sup(r) = \frac{sup\_count(r)}{N_L}$$

$$= \frac{\sum_{i=1}^{n_p} \sum_{j=1}^{|T(p_i)|} \left( \frac{sup\_count(p_i \wedge t_j \wedge r)}{sup\_count(p_i \wedge t_j)} \cdot sup\_count(p_i \wedge t_j) \right)}{N_L}$$

$$= \frac{\sum_{i=1}^{n_p} \sum_{j=1}^{|T(p_i)|} \left( \frac{sup(p_i \wedge t_j \wedge r)}{sup(p_i \wedge t_j)} \cdot sup\_count(p_i \wedge t_j) \right)}{N_L} \quad (4)$$

$$= \frac{\sum_{i=1}^{n_p} \sum_{j=1}^{|T(p_i)|} \left( conf(p_i \wedge t_j \Rightarrow r) \cdot sup\_count(p_i \wedge t_j) \right)}{N_L}$$

And in addition,

$$\sum_{i=1}^{n_p} \sum_{j=1}^{|T(p_i)|} \left( sup\_count(p_i \wedge t_j) \right) = N_L \quad (5)$$

Therefore,

$$sup(r) = \frac{\sum_{i=1}^{n_p} \sum_{j=1}^{|T(p_i)|} \left( conf(p_i \wedge t_j \Rightarrow r) \cdot sup\_count(p_i \wedge t_j) \right)}{\sum_{i=1}^{n_p} \sum_{j=1}^{|T(p_i)|} \left( sup\_count(p_i \wedge t_j) \right)} \quad (6)$$

So the support value $\sup(r)$ actually equals to the arithmetic mean values of all the confidence values in $\{conf(p_i \wedge t_j \Rightarrow r) | i=1,2,3,...,n_p, j=1,2,3,...,|T(p_i)|\}$.

Divide the both sides of Equ.(6) with $\sup(r)$ and we get:

$$\frac{\sum_{i=1}^{n_p}\sum_{j=1}^{|T(p_i)|}\left(\frac{conf(p_i \wedge t_j \Rightarrow r)}{\sup(r)} \cdot \sup\_count(p_i \wedge t_j)\right)}{\sum_{i=1}^{n_p}\sum_{j=1}^{|T(p_i)|}\left(\sup\_count(p_i \wedge t_j)\right)} = 1 \quad (7)$$

It means that when $\frac{conf(p_i \wedge t_j \Rightarrow r)}{\sup(r)} < 1$, it is pretty sure that in some other activity, say ($p_x \wedge t_y$, where $x \neq i$ and $y \neq j$), $\frac{conf(p_x \wedge t_y \Rightarrow r)}{\sup(r)} > 1$, and therefore, $conf(p_x \wedge t_y \Rightarrow r) > \sup(r) > conf(p_i \wedge t_j \Rightarrow r)$. Therefore, in a holistic view we claim that the resource $r_{17}$ is actually more suitable for some other work ($p_x \wedge t_y$). When the primary resource $r_4$ is unavailable, and if the administrator unwisely selects the secondary resource $r_{17}$ for $p_1 \wedge t_2$ from Rule 2, the conflict would occur when $p_x \wedge t_y$ requires $r_{17}$. In the context of data mining, the division equation $conf(LHS \Rightarrow RHS)/\sup(RHS)$ is named *lift* measure by definition[14, 23]:

$$\frac{conf(LHS \Rightarrow RHS)}{\sup(RHS)} = \frac{P(LHS,RHS)}{P(LHS)P(RHS)} \quad (8)$$
$$= lift(LHS \Rightarrow RHS)$$

*Lift* is a correlation measure used to find out uninteresting rules. A rule $LHS \Rightarrow RHS$ is negatively correlated if $lift(LHS \Rightarrow RHS) < 1$, else, it is positively correlated. In this paper, we annotate the negatively correlated rules and recommend them to the administrators as alternative resource candidates, along with the positive ones.

The negatively correlation annotation indicates that, although it is appropriate to assign the annotated resources to the workflow activity, it is better to keep them in reserve for their primary works.

The negatively-correlated-rule-annotation algorithm is as follows:

**Annotate negatively correlated resource allocation rules in the strong rules**

**Algorithm: Negative correlated association rules annotation.**
Annotate the rules with negative correlation.

**Input:**
- $S$, the candidate strong resource allocation rule set;

**Output:** $AR$, resource allocation rule with negative correlation annotation;

**Method:**
(1) **for each** resource allocation rule $rl \in S$ { //Scan $L$ for counts
(2) **if** $lift(rl) < 1$ **then**
(3)    annotate $rl$ as negative correlated; // Annotate the negative correlated rules
(4) **else add** $rl$ to $AR$;

### 4.3. Rules confidence ranking: sort rules by confidence

In association rule mining area, a major method to sort a collection of association rules is the most-confident selection method[24, 25]. The most-confident rule selection method always chooses the highest confidence among all the association rules whose support value is above the *min_sup* threshold. Hence, we use the confidence measure to sort the resulting rules to generate the resource allocation rules list for decision support.

When the PTR rules are generated, the rules are then divided into different sets by their LHS. Suppose that for a specific rule set with the LHS ($p_3 \wedge t_6$), the mined strong PTR rules are:

Rule 1: ProcessID=3 TaskID=6 ==> ResourceID=7   conf:(0.26) [support_count=426]
Rule 2: ProcessID=3 TaskID=6 ==> ResourceID=11  conf:(0.54) [support_count=885]
Rule 3: ProcessID=3 TaskID=6 ==> ResourceID=13  conf:(0.10) [support_count=164]

In this example, the confidence values of Rule 1, Rule 2, and Rule 3 are 0.26, 0.54 and 0.10, respectively. Then we get the ranked rule list by the confidence measure in descendant order:

Rule 2: ProcessID=3 TaskID=6 ==> ResourceID=11   conf:(0.54) [support_count=885]
Rule 1: ProcessID=3 TaskID=6 ==> ResourceID=7    conf:(0.26) [support_count=426]
Rule 3: ProcessID=3 TaskID=6 ==> ResourceID=13   conf:(0.10) [support_count=164]

With most-confident selection method, the system then automatically chooses the resource $r_{11}$ from rule 2 as default recommendation for the administrator. Note that in our approach, the system will also recommend the resources suggested by rest of the list, $r_7$ and $r_{13}$ (from Rule 1 and Rule 3) as alternatives. For N different test cases, let C be the number of correct predictions, then the resource prediction accuracy of the activity ($p_3 \wedge t_6$) is:

$$precision = \frac{C}{N} \qquad (6)$$

The rationale of most-confident selection method is that the testing data will share the same characteristics as the training data [25, 26]. Thus, if a rule has a high confidence in the training data, then this rule would also show a high accuracy in the testing data.

## 5. Experiment and evaluation

### 5.1. Experiment setup

Our work is based on the workflow history data from a PDM system named KM PDM (http://www.kmsoft.com.cn/Contents-119.aspx) deployed in a large electronic manufacturing enterprise[13] in Nanjing, China. We import the event data of 10 processes from the KM PDM database using SQL queries. Given the workflow log data, the first step is to clean the raw data. We filter out noise logs with no originators and those logs performed automatically or allocated to originators at design-time (The existence of these event logs will not help us in mining the run-time resource allocation rules). Finally, we get a log with 75934 items.

### 5.2. Training Data overview

Table 3 shows the execution frequency counts of the training log, each column of the table shows the task sequence number in the process, and each row corresponds to a process. As we can see, the columns of the table are the processes, and the rows represent the # of the tasks in processes, and the numbers in the cells are the frequency counts of the corresponding process and task. There are 10 processes and 141 tasks in the training dataset.

**Table 3** Process-task distribution in training data

| Task \ Process | 1 | 2 | 3 | 4 | 5 | 6 | 7 | 8 | 9 | 10 |
|---|---|---|---|---|---|---|---|---|---|---|
| 1 | 717 | 1020 | 281 | 380 | 609 | 642 | 206 | 655 | 755 | 1118 |
| 2 | 478 | 568 | 1109 | 720 | 561 | 879 | 589 | 869 | 608 | 248 |
| 3 | 335 | 284 | 764 | 798 | 562 | 208 | 253 | 777 | 922 | 183 |
| 4 | 240 | 786 | 671 | 335 | 502 | 535 | 567 | 173 | 776 | 398 |
| 5 | 278 | 182 | 722 | 483 | 811 | 616 | 718 | 730 | 310 | 642 |
| 6 | 715 | 197 | 370 | 690 | 424 | 421 | 677 | 690 | 262 | 976 |
| 7 | 207 | 275 | 1201 | 507 | 85 | 784 | 395 | 715 | 563 | 565 |
| 8 | 446 | 304 | 1189 | 441 | 296 | 792 | 628 | 953 | 837 | 619 |
| 9 | 258 | 741 | 613 | 269 | 773 | 1053 | 829 | 1048 | 314 | 204 |
| 10 | 644 | 226 | 281 | 467 | 0 | 166 | 178 | 160 | 376 | 461 |
| 11 | 1095 | 407 | 583 | 559 | 0 | 521 | 798 | 447 | 187 | 459 |
| 12 | 258 | 588 | 479 | 915 | 0 | 0 | 845 | 862 | 398 | 0 |
| 13 | 1168 | 524 | 429 | 319 | 0 | 0 | 813 | 0 | 718 | 0 |
| 14 | 810 | 377 | 798 | 0 | 0 | 0 | 828 | 0 | 307 | 0 |
| 15 | 985 | 456 | 0 | 0 | 0 | 0 | 0 | 0 | 318 | 0 |
| 16 | 957 | 1378 | 0 | 0 | 0 | 0 | 0 | 0 | 490 | 0 |

Fig. 4-6 demonstrates the basic properties of the workflow log in this paper. The X-axes are the # of the resources, activities and processes; the vertical axis represents the occurrence frequency or relative frequency.

Fig. 4 illustrates the occurrence frequencies of the resources in the training dataset.

Fig. 5 shows the frequency counts distribution of each activity. The vertical axis represents the perform times of each activity.

Fig. 6 illustrates the relative frequency distribution of the processes.

Note that in Fig. 5, we re-denote the workflow tasks with the term "activity" to illustrate the properties in 2-dimentional figures, and the id of the activities are:

$activity\_id = \max(task\_id) \bullet process\_id + task\_id$

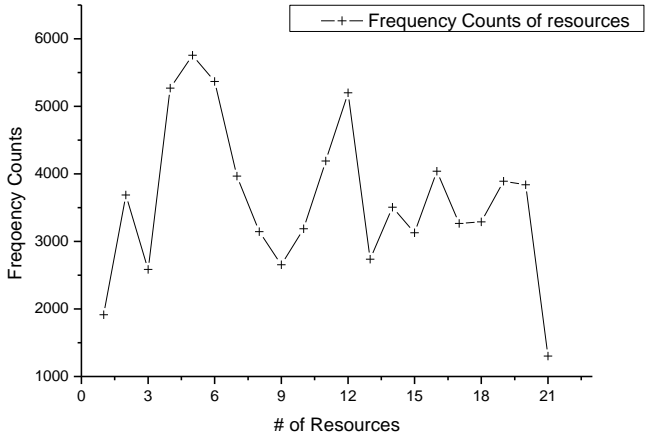

**Fig. 4** Frequency counts of resources of the training dataset

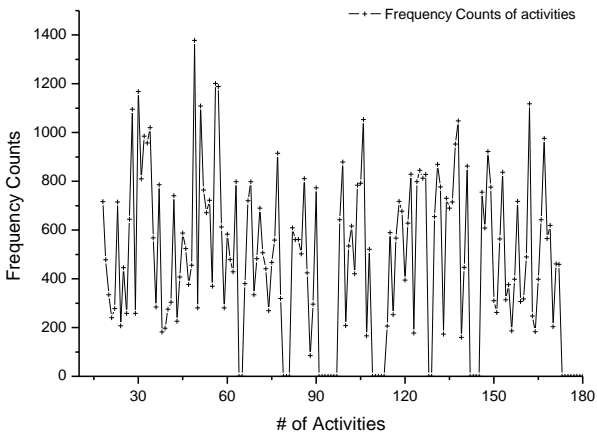

**Fig. 5** Frequency counts of activities of the training data

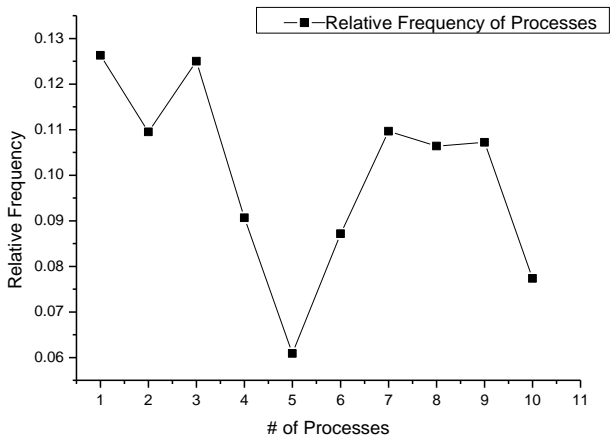

**Fig. 6** Process distribution in the training data

## 5.3. Parameters selection

The measure *min_sup* has a strong effect on the quality of the rules mined. On one hand, if *min_sup* is set too high, those possible rules that cannot satisfy the *min_sup* threshold but with high confidence may be excluded, and this directly affect the prediction accuracy of the rules. On the other hand, when *min_sup* is set too low, the mining process will be time-consuming [24, 25, 27]. Therefore, for the support value, we have to balance efficiency against quality. From our experiments, we observe that for our training set, once *min_sup* is lowered to 0.001%, the rules mined are more accurate than the classifier built by C4.5.

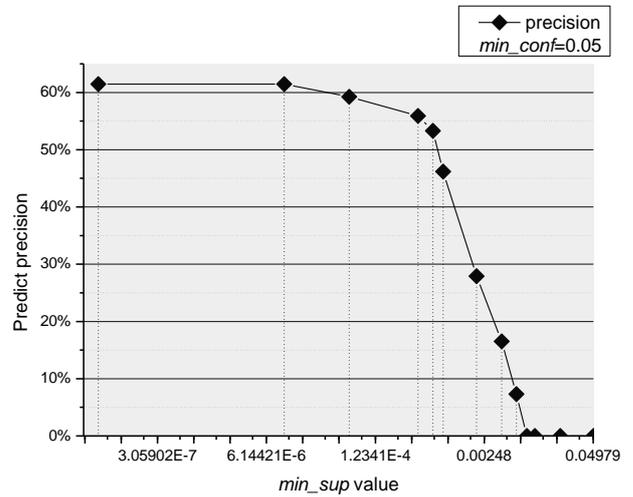

**Fig. 7** Overall accuracy under different *min_sup* thresholds

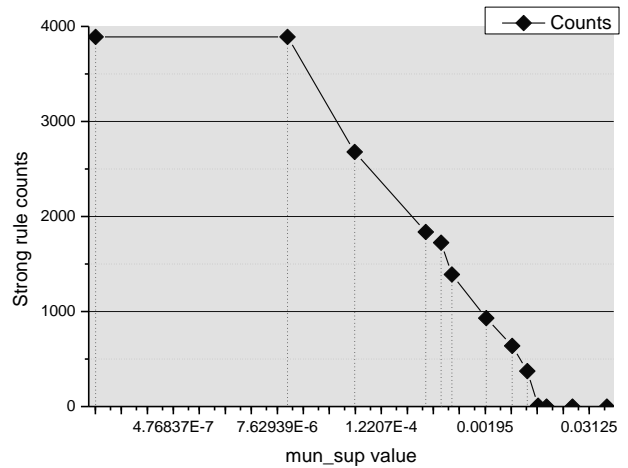

**Fig. 8** Number of strong rules under different *min_sup* thresholds

From Fig. 7 and Fig. 8, we can see that when *min_sup* threshold is lowered to 0.001%, the overall accuracy of the rules will go to the upper limit, 61.453%. Therefore, in this paper, we set *min_sup*=0.001%. We also set a limit of 20,000 on the total number of candidate rules in memory (including those dropped-off rules that do not satisfy either *min_sup* or *min_conf*).

## 5.4. Experiment results

After the preparations made above, we use the Apriori algorithm to generate association rules from the workflow log. Fig. 9 illustrates the large 3-itemsets $L(3)$ found in the proceeding of association rules mining:

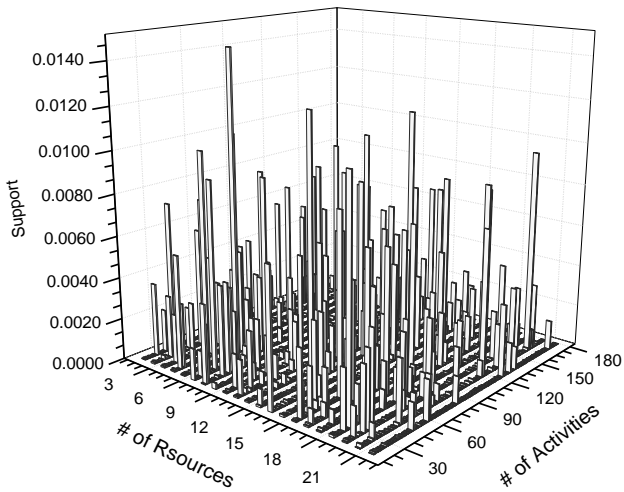

**Fig. 9** Large 3-itemsets $L(3)$ in the data cube found in the training set

After we get the large itemsets, we process the data with the 3-stage method referred in Section 4. For activity $act_{105}(p_6 \wedge t_9)$, we find in the log the rule list as:

**Stage 1. Generate the association rules:** With these large itemsets we can get the association rules above *min_sup* threshold and under the resource allocation constraint in Definition 3.

Rule 1:  process=6 task=9 1053 ==> resource=4 541   conf:(0.5138)
Rule 2:  process=6 task=9 1053 ==> resource=17 209 conf:(0.1985)
Rule 3:  process=6 task=9 1053 ==> resource=19 99   conf:(0.0940)
Rule 4:  process=6 task=9 1053 ==> resource=5 64    conf:(0.0608)
Rule 5:  process=6 task=9 1053 ==> resource=6 56    conf:(0.0532)
Rule 6:  process=6 task=9 1053 ==> resource=12 10   conf:(0.0095)
Rule 7:  process=6 task=9 1053 ==> resource=2  8    conf:(0.0076)
Rule 8:  process=6 task=9 1053 ==> resource=8  7    conf:(0.0066)
Rule 9:  process=6 task=9 1053 ==> resource=9  7    conf:(0.0066)
Rule 10:process=6 task=9 1053 ==> resource=20 7    conf:(0.0066)
Rule 11:process=6 task=9 1053 ==> resource=7  6    conf:(0.0057)
Rule 12:process=6 task=9 1053 ==> resource=10 6    conf:(0.0057)
Rule 13:process=6 task=9 1053 ==> resource=18 6    conf:(0.0057)

**Stage 2. Annotate the rules:** annotate the negatively correlated rules with mark "*".

**Stage 3. Sort the rules in precedence:** sort the rules with the confidence measure.

The final rule list can be:

Rule 1:     process=6 task=9 1053 ==> resource=4 541   conf:(0.5138) lift:7.4014

Rule 2:     process=6 task=9 1053 ==> resource=17 209 conf:(0.1985) lift:4.6118

Rule 3:     process=6 task=9 1053 ==> resource=19 99   conf:(0.0940) lift:1.8338

*Rule 4:    process=6 task=9 1053 ==> resource=5 64    conf:(0.0608) lift:0.8017

*Rule 5:    process=6 task=9 1053 ==> resource=6 56    conf:(0.0532) lift:0.7523

*Rule 6:    process=6 task=9 1053 ==> resource=12 10   conf:(0.0095) lift:0.1387

*Rule 7:    process=6 task=9 1053 ==> resource=2  8    conf:(0.0076) lift:0.1565

*Rule 8:    process=6 task=9 1053 ==> resource=8  7    conf:(0.0066) lift:0.1605

*Rule 9:    process=6 task=9 1053 ==> resource=9  7    conf:(0.0066) lift:0.1901

*Rule 10:   process=6 task=9 1053 ==> resource=20 7    conf:(0.0066) lift:0.1315

*Rule 11:   process=6 task=9 1053 ==> resource=7  6    conf:(0.0057) lift:0.1091

*Rule 12:   process=6 task=9 1053 ==> resource=10 6    conf:(0.0057) lift:0.1357

*Rule 13:   process=6 task=9 1053 ==> resource=18 6    conf:(0.0057) lift:0.1315

According to the rules above, the system predict resource $r_4$ as the originator of the activity, $act_{105}(p_6 \wedge t_9)$. Fig. 10 shows the prediction accuracy of all the activities using the most-confidence selection method:

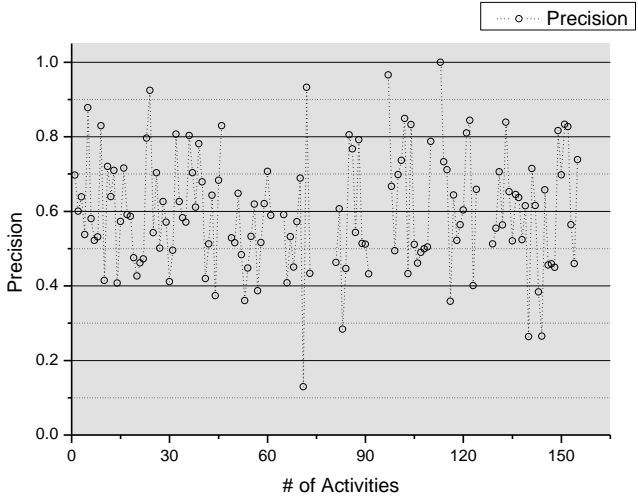

**Fig. 10** Accuracy of different activities

In the most confident selection method discussed in section 4.3, resource $r_4$ is the default originator for $act_{105}(p_6 \wedge t_9)$, and can be also viewed as a class label. Thus we can get a subset of most confident resource allocation rules, and build a classifier for each workflow activity. We make some comparison experiments between this Apriori-based classifier and the classification algorithms in [2, 28] using the data mining tool WEKA[29]. For the Apriori-based classifier[25], we set the parameters as: min_conf=0.05, min_sup=0.0001%, and rules number upper limit=10000; the others are four classification algorithms applied in Refs. [2, 28]: SVM, C4.5, ID3 and Naïve Bayes, considering the characteristics and variation of the training dataset, we set the test mode as 10-fold cross-validation.

Table 4 lists the number of correct predictions, mining time, and overall prediction accuracy of the algorithms.

**Table 4** Overall prediction accuracy of different methods

| Methods | Correct predictions counts | Time elapsed(s) | Overall prediction accuracy (%) |
| --- | --- | --- | --- |
| Apriori | 46663 | 20 | 61.452 |
| SVM | 46643 | 9268 | 61.426 |
| C4.5 | 46656 | 17 | 61.443 |
| ID3 | 45285 | 14 | 59.637 |
| Naïve Bayes | 28920 | 9 | 38.086 |

From Table 4, we conclude that except Naïve Bayes, other algorithms achieve an overall accuracy about 60%, and the values are very close. Naïve Bayes performs best in mining time, and the training/testing time of SVM is extremely long.

The performance of the proposed classifier based on the most confident selection method is reasonable compared with those in [2, 28]. However, the overall prediction accuracy of around 60% also implies that about 40% of all the system-assigned workflow activities need manual reassignments. Therefore, the rules with highest accuracy to the testing data are not always the best choice. Take $act_{105}(p_6 \wedge t_9)$ for example, Rule 1 is of a confidence 51%, and the sum of top 3 positively correlated rules reaches up to 80.63%.

Therefore, instead of suggesting one best prediction for each class of workflow activities, the system also recommend other strong resource allocation rules to the workflow administrators as candidates: when the resources with high confidence are unavailable at the moment, the remaining candidates (including the annotated resources) in the list could be the alternatives.

In addition, with the assistance of the negatively correlated rules annotations, the administrators can have a holistic view of the resources' work priorities. They can make assignments following the positively correlated rules in priority, and turn to negatively correlated ones only when all the prior resources are heavily occupied.

5.5. Special events and further discussion

Note that in Fig. 10, there are weak-predicated cases like the activity #71 (12.9%), #83 (28.4%) and #140 (26.4%) etc., in fact, these activities have some common features. Through further analysis, we find the reasons as follows: firstly, the task is probably of relatively small number of training samples, like #71 and #140 (refer to Fig. 5), the inadequacy of training samples leads to a weak prediction. Another reason is, each work has been evenly assigned to many resources (like the activity #83, usually with more than 5 actors), therefore, there is actually no strict No. 1 actor for these activities and it usually does not matter which resource to reform the task.

In the proposed approach, the rules of different forms from the PTR rules are eliminated from the resource allocation recommendation list. However, we find some of these intermediate products of interestingness. Following are two examples:

Let us see a strong RP rule (resource to process): $r_{17} \rightarrow p_3, conf = 0.73$, this rule implies that 73% of the work of resource $r_{17}$ locates in process $p_3$, so we may infer that resource $r_{17}$ is skilled in the tasks in $p_3$.

Besides, the TP rule (task to process) $t_{16} \rightarrow p_2, conf = 0.49$ indicates that about half of the tasks $t_{16}$ are in process $p_2$, which just conforms with the statistic results in Table 3.

These "by-products" have no distinct contribution to workflow resource allocation, but still of reference values to the workflow administrators.

## 6. Related work

The workflow technology provides a broad support to manage the works running in information systems. Such generic information systems that are configured on the basis of process models are referred to as process-aware information systems (PAISs, e.g., workflow management systems, ERP systems, CRM systems, PDM systems), and are now widely used in manufacturing enterprises.

Nowadays, a hot topic in the workflow context is to find and use the knowledge in the workflow management procedures[30-34]. Current PAISs usually record all kinds of events, the omnipresence of event logs in PAISs is a motivator of process mining. Process mining is a state-of-art technology in discovering useful information (e.g., knowledge of process control flow or organizational structures) from event logs [15, 35]. For different mining perspectives, the result varies: the control flow mining is in the process perspective; and so are the organizational architecture and relationships of the workflow systems.

As far as we know, despite of the great efforts spent on the control flow and data aspect of workflow, the organizational aspect of processes have been often neglected. To date, there has been a relatively small body of researches in workflow resource allocation. However, in order to fully understand workflow, it is very important to find the relationship between the processes and the resources[2], e.g. by whom the activities should be performed.

The target of the allocation of the workflow tasks to resource is to find the logic between the workflow process, the activity, and the resources. In the organizational aspect of process mining, according to [15, 35], there are four measures: measures based on (possible) causality, measures based on joint cases, measures based on joint activities, measures based on special event types. To the best of our knowledge, related researches so far have made classification as a popular choice to discover resource allocation knowledge from the workflow logs, and most of the recent research activities in resource allocation fall into process mining[9] in the organizational perspective, namely organizational mining, related work is as follows:

Ref. [2] discussed an approach to semi-automating the run-time staff assignment in workflow management systems. In order to reduce the amount of manual staff assignments, Y. Liu et al. apply the machine learning technology to the workflow event log from three enterprises to learn the kinds of activities that each staff undertakes. In Ref. [1], Ly et al. shows that the task of mining staff assignment rules using history data and organizational information can be considered as a inductive learning problem, and they adapt a decision learning approach to derive staff assignment rules.

In Ref. [28], Rinderle and van der Aalst develop a framework for the complete life-cycle support for staff assignment rules. Ref. [36] uses Hidden Markov Models to allocate the proficient staffs for a whole business process based on the workflow event log. In Ref. [37], Andrzejak et al. present a closed-loop workflow framework in control flow aspect, that implements a general closed control loop of planning – execution – result validation – re-planning, and generates workflows.

In our previous work in Refs.[38], we present a closed-loop workflow management framework: we apply a statistic approach to derive resource allocation knowledge from the workflow log to assist assigning the resources for the upcoming workflow tasks. The mined staff assignment information is then feedback to construct a closed-loop in workflow resource management.

As is shown in the previous works mentioned above, researchers concentrate on finding the suggestion with the highest accuracy for a class of workflow activities. However, these best suggestions may not always be the best choices, sometimes the best prediction comes from the administrator's judgments, according to the real-time situation of the system. Our approach is different from that of Refs. [1, 2, 28, 36] in that we not only give the best prediction for each activity, but also give a recommendation of candidate predictions. Such strategy makes it an easier way for the administrators to reach for applicable alternatives when the default prediction fails.

The idea of automatic resource allocation can also be found in the literature on advanced manufacturing technologies, typically, the Advanced Planning and Scheduling (APS)[39]. APS refers to a manufacturing process management (MPM) by which the production resources (including materials and production capacity) are optimally allocated to meet the manufacturing demands. In Ref. [40], Stadtler discusses the essence of SCM and advanced planning in the form of two conceptual frameworks: The house of SCM and the supply chain planning matrix. In Ref. [39], Lee et al. present a model for advanced planning and scheduling (APS) that requires an absolute due date with outsourcing in a manufacturing supply chain. The proposed model considers alternative process plans for job types, with precedence constraints for job operations. Another research about advanced resource planning is proposed by Vandaele et al.[41, 42]. They propose a decision support module for the manufacturing planning and control system called advanced resource planning (ARP). The ARP module provides a parameter-setting process, with the ultimate goal of yielding realistic information about production lead times for scheduling purposes, sales and marketing, strategic and operational decision making, and suppliers and customers.

## 7. Conclusions and future work

We have presented a decision-making approach using data mining technology to make recommendations to workflow initiators. In the closed-loop workflow resource allocation framework, the association rules mining algorithms are applied to the workflow event log for mining resource allocation rules. Our current research is oriented towards developing more productive WfMS in resource management along the following lines: (a) implementing the proposed framework in a web-based architecture, (b) association rules mining to generate strong resource allocation rules, (c) using the negative correlation measures to annotate the negative correlated rules, (d) ranking the rules to make decision support for resource allocation.

To illustrate, we make some comparison experiments on the log data distract from a manufacturing enterprise, experiment results show an overall accuracy of over 50%, and we made a comparison between the presented approach and the classification algorithms and analyzed their performances. Feasibility evaluation via a case study suggests that the proposed approach would be useful in supporting workflow resource allocation.

Then we discuss the advantages and limitations of the method. Along with the administrators' awareness of the workload of the resources, and professional knowledge to different product design tasks, our approach can well handle most of the resource allocation problems in PAISs.

Our future work includes two main parts: (1) compare some other machine learning approaches like inductive learning programming (ILP) with our present method to find some more efficient and effective approaches. (2) find the resource allocation rules from different organizational levels and dimensions (e.g. the roles and the organizational units).

**Acknowledgement**

This work is funded by certain ministry of China under Grant 51318010103 and Grant 9140A18010111JW0602. We are also very grateful to anonymous referees for their valuable comments.